\documentclass{iopart}
\usepackage{iopams,amsfonts,amssymb,amsthm} 
\usepackage{geometry}                
\usepackage{graphicx}
\usepackage{rawfonts}
\usepackage{amssymb}
\usepackage{epstopdf}
\usepackage[usenames,dvipsnames,svgnames]{xcolor}
\input prepictex.tex
\input pictex.tex
\input postpictex.tex

\begin{document}

\newtheorem{theo}{Theorem}
\newtheorem{lemm}{Lemma}
\newtheorem{corr}{Corollary}
\newtheorem{defn}{Definition}
\newtheorem{remark}{Remark}

\def\Pr{{\it Proof:  }}
\def\qed{$\Box$}

\article{}{Self-avoiding walks subject to a force} 
\author{E J Janse van Rensburg$^1$\footnote[1]{\texttt{rensburg@yorku.ca}}
 and S G Whittington$^2$\footnote[1]{\texttt{swhittin@chem.utoronto.ca}}}
\address{$^1$ Department of Mathematics, York University, Toronto, Canada}
\address{$^2$ Department of Chemistry, University of Toronto, Toronto, Canada}
\begin{abstract}
We prove some theorems about self-avoiding walks attached to an
impenetrable surface (\emph{i.e.} positive walks)
and subject to a force.  Specifically we show 
the force dependence of the free energy is identical when the force is applied
at the last vertex or at the top (confining) plane.  We discuss the relevance of this result to 
numerical results and to a recent result about convergence rates when the walk 
is being pushed towards the surface.

\end{abstract}

\pacs{82.35.Lr,82.35.Gh,61.25.Hq}
\ams{82B41, 82B80, 65C05}
\submitto{J Phys A}
\maketitle

\section{Introduction}
\label{sec:Introduction}
The introduction of micro-manipulation techniques such as atomic force 
microscopy (AFM) and optical tweezers \cite{Haupt1999,Zhang2003} has led to 
an interest in the theoretical description of polymer molecules subject to a force.
If we are interested in linear polymers then the natural model is a self-avoiding
walk \cite{Hammersley1957,MadrasSlade}.  
Consider the $d$-dimensional hypercubic lattice, $Z^d$, and attach
the obvious coordinate system $(x_1,x_2,\ldots x_d)$ so that each vertex of the 
lattice has integer coordinates.  If we are interested in polymers interacting with a 
surface we can take the hyperplane $x_d=0$ as the relevant surface and consider
self-avoiding walks starting at the origin and with no vertices having negative
$x_d$-coordinate.  These are called \emph{positive walks}.

Suppose that $c_n^+(v,h)$ is the number of $n$-edge positive walks 
with $v+1$ vertices in $x_d=0$ and with the $x_d$-coordinate of their
last vertex equal to $h$.  Define the partition function as 
\begin{equation}
C^+_n(a,y) = \sum_{v,h} c_n^+(v,h)a^vy^h
\end{equation}
where $a=e^{-\epsilon /kT}$ and $y=e^{f/kT}$.  $\epsilon$ is the energy associated with a 
vertex in the surface, $f$ is the applied force, $k$ is Boltzmann's constant and $T$
is the absolute temperature.
This is  a model for polymers interacting with the surface so that the polymer
can be adsorbed, with a force applied normal to the surface to pull the polymer off the surface.
There are some rigorous results about this problem \cite{Guttmann2014,Rensburg2013}
(see section 9.7 in reference \cite{Rensburg2015}),
as well as several numerical studies either by Monte Carlo methods \cite{Krawczyk2005}
or by exact enumeration and series analysis \cite{Guttmann2014,Mishra2005}.  See
\cite{IoffeVelenik,Skvortsov2009,Binder2012} for related work.

The problem has independent interest if $a=1$ so that there is no interaction
with the surface (except that the surface is impenetrable) \cite{Beaton2015,Rensburg2009}.
In particular Beaton \cite{Beaton2015} has shown that the walk is ballistic for any 
$f>0$.  See also \cite{IoffeVelenik}.
There are some results about the related problem of polygons pulled away from a surface
as a model of ring polymers \cite{Rensburg2008a,Rensburg2008b},

\begin{figure}[h]
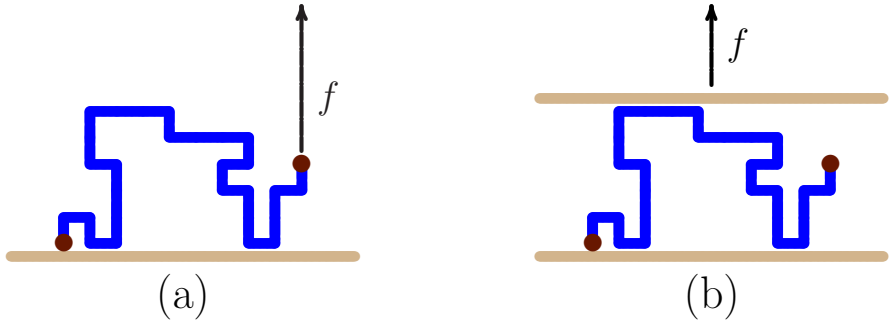

\centering
\input figure1.tex
\caption{(a) A positive walk pulled at its endpoint in the vertical direction from a surface.
(b) A positive walk pulled at its highest vertices in the vertical direction.}   
\label{figure1}
\end{figure}

In the treatments of self-avoiding walks described above
\cite{Beaton2015,Guttmann2014,Rensburg2013,Krawczyk2005,Mishra2005}
the force is applied at the last vertex of the walk, as in figure \ref{figure1}(a).
What happens if the force is applied 
in some other way?   
In an AFM experiment the monomer in contact with the tip
will not always be the last monomer in the polymer and it is interesting to enquire how robust
the results are.  In a recent paper \cite{GuttmannLawler} the force is applied differently.  
Their idea is to apply the force in the plane containing the vertices of the walk that are furthest
from the surface (see figure \ref{figure1}(b)).  Suppose that $c_n(v,s)$ is the number of $n$-edge positive
walks with $v+1$ vertices in the surface and with span in the $x_d$-direction equal to $s$.
The corresponding partition function is 
\begin{equation}
C_n(a,y) = \sum_{v,s} c_n(v,s) a^vy^s
\end{equation}
When $a=1$ (no surface attraction) the problem has been considered in
\cite{GuttmannLawler,Rensburg2009}.  

Instead of pulling a walk away from the surface one can push the walk towards
the surface ($f<0$ or $y<1$) \cite{GuttmannLawler,Rensburg2009}.  Having the force applied
at the last vertex or in the confining plane seems then to be very different and, for small
$n$, this is apparent from Monte Carlo data \cite{Rensburg2009}.  In \cite{GuttmannLawler}
the authors consider the finite $n$ behaviour in the latter case and they find unexpected 
subdominant correction terms.  These are probably not present in the first case (with the 
force applied at the last vertex) \cite{GuttmannLawler,Guttmann2014}.  In this paper we compare and contrast the behaviour with these 
two ways of applying the force.  We consider both pushing the walk towards the 
surface ($y<1$) and pulling away from the surface ($y>1)$.

\section{Bridges subject to a force}
\label{sec:bridges}

We shall be concerned with the situation where there is no attractive 
interaction with the surface.  We define the two generating functions
\begin{equation}
C^+(y,z) = \sum_n C_n^+(1,y) z^n, \quad  C(y,z) = \sum_n C_n(1,y) z^n.
\end{equation}
We define a \emph{bridge} as a positive walk with the extra conditions that
\begin{enumerate}
\item 
the first edge is in the $x_d$-direction, and the walk does not return to the 
hyperplane $x_d=0$;
\item
the $x_d$-coordinate of the last vertex is at least as large as that of any
other vertex.
\end{enumerate}
Let $b_n(h)$ be the number of $n$-edge bridges with the $x_d$-coordinate of the last vertex being 
$h$.  Define the generating function
\begin{equation}
B(y,z) = \sum_{h,n} b_n(h) y^h z^n.
\end{equation}
Define the slab $S_w$ to be the set of lattice vertices with $x_d$-coordinate satisfying
$0 \le x_d \le w$.   Define the generating function of bridges that span $S_w$ as
\begin{equation}
B_w(z) = \sum_{n=w}^{\infty} b_n(w) z^n.
\end{equation}

\begin{lemm}
$B_w(z)$ is singular at $z=z_w$ where $z_w \ge z_{w+1} \ge 1/\mu$ and 
$\inf_w z_w = 1/\mu$.  
\end{lemm}
\Pr
If we delete the first edge of a bridge with $n+1$ edges in a slab with span $w+1$, 
translate through unit distance in the negative
$x_d$-direction and decrease the width of the slab by unity we obtain a walk
with $n$ edges in a slab of width $w$.  Clearly $b_{n+1}(w+1) \le c_n(1,w)$.  
Consider positive walks with $n$ edges confined to a slab
of width $w$. Unfold each walk in the $x_1$-direction
\cite{HammersleyWelsh}.  At most $e^{O(\sqrt n)}$ walks give rise to the same 
unfolded walk \cite{HammersleyWelsh}.  Suppose that the last vertex of the unfolded walk has 
$x_d$-coordinate equal to $w-q+1$.  Add an edge in the positive $x_1$-direction and then add 
$q-1$ edges in the positive $x_d$-direction so that the final vertex is in $x_d=w$.  Convert
this to a bridge with span $w+1$, unfolded in the $x_1$-direction, with $n+q+1$ 
edges by adding an additional edge at the beginning of the walk.  Therefore 
$c_n(1,w) \le b_{n+q+1}(w+1) e^{O(\sqrt n)}$.  These two inequalities imply that the free energy of 
bridges in a slab of width $w+1$ is equal to that of walks in a slab of width $w$.  Since it 
is known that the free energy of walks is strictly increasing in $w$ and its limit is 
$\log \mu$ \cite{HammersleyWhittington} this proves the lemma.
\qed

\begin{theo}
The radius of convergence, $z_c^B(y)$, of $B(y,z)$ is equal to 
$1/\mu$ for all $y \le 1$, where $\mu$ is the growth constant of self-avoiding
walks.
\label{theo:bridgepush}
\end{theo}
\Pr
Since $b_n(h) \le c_n^+(h)$ it is clear that $B(y,z) \le C^+(y,z)$.  We know that the 
radius of convergence of $C^+(y,z)$ is $z_c^+(y) = 1/\mu$ for $y\le 1$ \cite{Rensburg2013}
where $\mu$ is the growth constant of self-avoiding walks.  Hence the radius of 
convergence of bridges, $z_c^B(y)$ is bounded by
\begin{equation}
z_c^B(y) \ge z_c^+(y) = 1/\mu, \quad \forall y \le 1.
\end{equation}
Now 
\begin{equation}
B(y,z) = \sum_n \sum_w b_n(w) y^w z^n \ge y^w \sum_{n=w}^{\infty} b_n(w) z^n = y^w B_w(z)
\end{equation}
for any $w >0$ and $y \le 1$.  Hence $z_c^B(y) \le z_w $ for all $w$ and 
\begin{equation}
z_c^B(y) \le \inf_w z_w = 1/\mu
\end{equation}
for all $y \le 1$.  Hence $z_c^B = 1/\mu$ for all $y \le 1$. \qed

\section{Self-avoiding walks subject to a force}
\label{sec:saws}
We now turn to the problem of positive walks confined between two parallel 
planes with the planes being pushed together.  The following Theorem follows easily
from Theorem \ref{theo:bridgepush}.

\begin{theo}
The radius of convergence, $z_c(y)$, of the generating function
$C(y,z)$ is equal to $1/\mu$ for all $y \le 1$.
\label{theo:walkpush}
\end{theo}
\Pr
Clearly $C(y,z) \le C(1,z)$ for all $y \le 1$ by monotonicity.  But $C(1,z)$ is
the generating function of positive walks and has radius of convergence equal to 
$1/\mu$ \cite{Whittington} so the radius of convergence of 
$C(y,z)$,  $z_c(y)$, is bounded by $z_c(y) \ge 1/\mu$ for $y < 1$. 
By inclusion $C(y,z) \ge B(y,z)$ so $z_c(y) \le z_c^B(y)$.  By Theorem
\ref{theo:bridgepush} we have $ z_c^B(y) = 1/\mu$ for $y \le 1$ and this proves the 
Theorem.
\qed

If the walk is being pulled away from the surface, so that $y > 1$, we have the following
theorem.

\begin{theo}
The radii of convergence of the generating functions $C(y,z)$, $C^+(y,z)$ and $B(y,z)$
are all equal when $y \ge 1$.
\label{theo:walkpull}
\end{theo}
\Pr
Since every bridge is also a walk counted by $C^+(y,z)$ and by $C(y,z)$ we have
$B(y,z) \le C^+(y,z)$ and $B(y,z) \le C(y,z)$ by inclusion.  All positive walks are 
counted both by $C^+(y,z)$ and by $C(y,z)$ and the span of a walk is always at least as 
large as the height of its last vertex ($s \ge h$).  Hence, when $y > 1$, each walk 
receives at least as large a weight in $C(y,z)$ as in $C^+(y,z)$ so $C^+(y,z) \le 
C(y,z)$ when $y > 1$.  Of course $C^+(1,z)=C(1,z)$.  Hence
\begin{equation}
B(y,z) \le C^+(y,z) \le C(y,z), \quad y>1
\end{equation}
and therefore their radii of convergence are related by
\begin{equation}
z_c^B(y) \ge z_c^+(y) \ge z_c(y).
\label{eqn:zcinequality}
\end{equation}
Each walk counted by $C(y,z)$ can be converted to a bridge by unfolding in the 
$x_d$-direction and at most $e^{O(\sqrt{n})}$ such walks give the same 
bridge \cite{HammersleyWelsh}.  Moreover the span can not decrease 
in the unfolding operation so, for $y > 1$, $C_n(1,y) \le e^{O(\sqrt{n})} B_n(y)$.
This implies that $z_c(y) \ge z_c^B(y)$.  This, together with (\ref{eqn:zcinequality}),
proves the Theorem.
\qed

\section{Discussion}
\label{sec:discussion}

The theorems proved in Section \ref{sec:saws} establish that the force dependence of the 
free energy is identical when the walk is pulled or pushed at its last vertex and at the 
top (confining) plane.   In particular, the critical value $y_c=1$ (see reference
\cite{Beaton2015}) is the same for the two modes
of pulling.  These theorems are interesting in view of the results in
\cite{GuttmannLawler} and \cite{Rensburg2009}. 
In \cite{GuttmannLawler} the authors give convincing arguments and numerical evidence that,
in two dimensions when $y < 1$, the 
partition function $C_n(1,y)$ behaves asymptotically as
\begin{equation}
C_n(1,y) \sim \mbox{const} \times n^{3/16} \exp [-\mbox{const} \times u^{4/7} n^{3/7}] \mu^n,
\end{equation}
where $u = - \log y$.  The sub-exponential term $e^{n^{3/7}}$ leads to slow convergence
to the infinite $n$ behaviour.  See for instance \cite{Guttmannseries}.

In \cite{Rensburg2009}
the free energy as a function of the force is estimated numerically in three dimensions for both
modes of force application.  The results clearly support the implication of 
Theorem \ref{theo:walkpull}, and the free energies are very similar even for modest 
values of $n$.  When the walk is being pushed towards the surface ($y < 1$) the numerical 
results in \cite{Rensburg2009} show that there are large differences in the two 
free energies for values of $n$ as large as 4000.  The numerical results coupled with Theorem \ref{theo:walkpush} imply slow convergence to the limiting free energy 
when $y < 1$, and this is exactly the prediction of \cite{GuttmannLawler} for $d=2$.

\begin{figure}
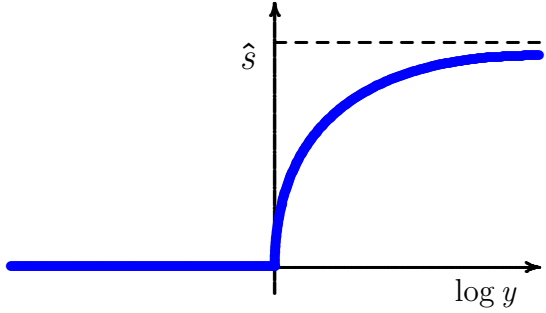

\centering
\input figure2.tex
\caption{The limiting mean scaled span $\widehat{s}$ as a function of $\log y$.  The curve is identical
for the two modes of pulling illustrated in figure \ref{figure1}.  For $y< 1$, $\widehat{s}=0$,
and for $y>1$ it is positive and asymptotic to $1$.  It is not known that this curve is continuous
at $y=1$.}
\label{figure2}
\end{figure}

For walks pushed or pulled in their confining plane we can write the average 
span, scaled by the number of edges,  as 
\begin{equation}
\frac{\langle s \rangle}{n} =\frac{1}{n} \frac{\sum_s  s c_n(s) y^s }{\sum_s c_n(s) y^s}= 
\frac{1}{n} \frac{\partial  \log C_n(1,y)}
{\partial  \log y}
\end{equation}
where $c_n(s)$ is the number of $n$-edge positive walks with span $s$. Taking the $n \to 
\infty$ limit gives 
\begin{equation}
\widehat{s}=\lim_{n\to\infty} \frac{\langle s \rangle}{n} = \frac{\partial 
[\lim_{n\to\infty} n^{-1} \log C_n(1,y)]}{\partial \log y}
\end{equation}
where we have used convexity \cite{Rensburg2009} to justify the interchange of the 
order of the limit and the derivative.  When $y < 1$ this limiting reduced span is zero, by Theorem
\ref{theo:walkpush}, so that 
the average span $\langle s \rangle = o(n)$ when $y<1$.  When $y>1$ the function
$\widehat{s}$ is positive and the $y$-dependence is sketched in figure \ref{figure2}.  Notice
that $\widehat{s}$ is asymptotic to $1$ in the large $y$ limit.  It is not known that $\widehat{s}$
is continuous at $y=1$.

\section*{Acknowledgement}
This research was partially supported by NSERC of Canada.

\end{document}